\documentclass[preprint]{aastex}

\shorttitle{QUASILINEAR PARTICLE DRIFT}
\shortauthors{Stawicki}
\begin{document}

\title{QUASILINEAR DRIFT OF COSMIC RAYS IN WEAK TURBULENT
  ELECTROMAGNETIC FIELDS}
\author{Olaf Stawicki}
\affil{Unit for Space Physics, North-West University, Potchefstroom
  Campus, Private Bag X6001, Potchefstroom 2520, South Africa}
\email{fskos@puk.ac.za}

\begin{abstract}
A general quasilinear transport parameter for particle drift in
arbitrary turbulence geometry is presented. The new drift
coefficient is solely characterized by a nonresonant term and is evaluated for slab and two-dimensional turbulence geometry. The calculations presented 
here demonstrate that fluctuating electric fields are a key quantity for understanding quasilinear particle drift in slab geometry. It is shown that particle drift 
does not exist in unpolarized and purely magnetic slab fluctuations. This is in stark contrast to previous models, which are
restricted to slab geometry and the field line random walk limit. The evaluation of the general transport parameter for two-dimensional
turbulence geometry, presented here for the first time for dynamical magnetic turbulence, results in a drift coefficient valid for a
magnetic power spectrum and turbulence decay rate varying arbitrarily in wavenumber. For a two-component, slab/two-dimensional turbulence model,
numerical calculations are presented. The new quasilinear drift, induced by the magnetic perturbations, is compared with a standard drift
expression related to the curvature and gradient of an unperturbed heliospheric background magnetic field. The considerations presented here offer a
solid ground and natural explanation for the hitherto puzzling observation that drift models often describe observations much better
when drift effects are reduced.
\end{abstract}


\section{INTRODUCTION AND MOTIVATION}
\label{sec:intro}

Particle transport in random magnetic fields plays a key role in space physics and astrophysics. The knowledge of turbulence properties
is crucial for understanding (anisotropic) particle diffusion in a collisionless, turbulent and magnetized plasma such as the solar wind
or the interstellar medium.  Of particular interest are processes governing particle diffusion across an ambient magnetic field, and
perpendicular diffusion and particle drifts appear to be the primary mechanisms. Transport coefficients describing these two processes are
thus key quantities for several heliospheric and astrophysical settings such as solar modulation of cosmic rays and diffusive shock
acceleration, e.g., at the heliospheric shock of termination or at shocks triggered by supernovae.

It is generally accepted that the modulation of cosmic rays in the heliosphere is described adequately by Parker's equation
\citep{parker65} taking into account (anisotropic) spatial diffusion of cosmic rays by the diffusion tensor ${\mathsf K}$. This
tensor describes the interaction of cosmic rays with the turbulence. In a local orthogonal coordinate system with one axis along the direction
of the mean magnetic field, it takes on its simplest form and reads
\begin{equation}
{\mathsf K}=
\pmatrix{\kappa_{\perp,1} &  \kappa_A  & 0
     \cr   -\kappa_A     &  \kappa_{\perp,2}  &     0
     \cr  0 &  0  & \kappa_{\|}
}.
\label{eq:crdc}
\end{equation}
The diagonal elements describe diffusion along ($\kappa_{\|}$) and perpendicular ($\kappa_{\perp,1}$, $\kappa_{\perp,2}$) to the mean
magnetic field. The off-diagonal, antisymmetric entry $\kappa_A$ is related to effects of curvature and gradient drift in a nonhomogeneous
heliospheric magnetic field (HMF). For an unmodified Parker spiral \citep{par58}, $\kappa_A$ reads \citep{jok_etal77} 
\begin{equation}
\kappa_{A}=\frac{vR_L}{3},
\label{eq:largedrift}
\end{equation}
where $v$ and $R_L$ denote the speed and the Larmor radius of a particle, respectively. The derivation of
equation (\ref{eq:largedrift}) is based on the assumption that the HMF is undisturbed, i.e., electromagnetic turbulent fields are not taken into account. 

The importance of large-scale drifts of cosmic rays in the HMF was first recognized in the late seventies by \citet{jok_etal77} and
subsequently taken into account, via equation (\ref{eq:largedrift}), in enumerable numerical studies \citep[e.g.,][]{pot_bur90,web_etal90,pot_etal93,bur_hat98}. While
being indispensable for a description of multi-dimensional large-scale modulation, it has been demonstrated that drift effects do not have to
be included in all cases to reproduce observations \citep[see, e.g.,][]{rei_etal93,roux_fic97}. 

It has been pointed out in several numerical studies that the effect on modulation, as described by equation (\ref{eq:largedrift}), is too large if the numerical results 
are compared with observations \citep[e.g.,][]{potgieteretal1987,potgieteretal1989,pot_bur90,web_etal90}. To obtain a better agreement between simulations and 
observations, it was suggested to use a reduced amount of drift effects at low and intermediate particle energies \citep[][]{potgieteretal1987,potgieteretal1989,bur90}. 
In spite of its importance and the pressing need of rigorous theories, only one model has been developed in the past being able to explain a possible reduction of
the drift at intermediate energies \citep{bie_mat97}. In brief, they obtain the expression
\begin{equation}
\kappa_{A}=\frac{vR_L}{3}\frac{(\tau\Omega)^2}{1+(\tau\Omega)^2}
\label{eq:bams}
\end{equation}
with $\Omega$ being the relativistic particle gyrofrequency. The timescale $\tau$ is associated with the decorrelation of a particle
trajectory from an unperturbed helical orbit. For their calculations, \citet{bie_mat97} assume slab turbulence geometry and argue that
nonresonant field line random walk (FLRW) is the major agent for the decorrelation of particle trajectories. Based on this, they demonstrate
that drift effects can be suppressed at low and intermediate particle energies. Another model, developed earlier by \citet{for_etal74} on the basis of
quasilinear theory (QLT) and FLRW in slab geometry, is formally based on the same expression for $\kappa_A$, equation (\ref{eq:bams}). It has been
shown, however, that this model fails in explaining suppressed drift effects, since their approach is valid only for cosmic ray
energies greater than $\sim 3$ GeV at Earth \citep[][]{bie_mat97}. Besides being valid for slab geometry only, both models take into account
purely magnetic fluctuations and were developed for static turbulence only.  Finally, for completeness, it should be mentioned that 
equation (\ref{eq:bams}) is formally the same as for hard-sphere scattering in a magnetized plasma \citep{gleeson1969}, where $\tau$ is then the scattering time. 
Furthermore, a similar structure was found for collisional particle transport parameter in a thermal equilibrium plasma \citep{balescuetal1994} and 
expressions for coefficients being of similar forms were already considered in the mid 1960s \citep[cf., e.g.,][and references therein]{toptygin1985}.

Upon looking into the literature, one becomes aware of the fact that a rigorous, quasilinear treatment of particle drift in an
electromagnetic plasma wave turbulence with arbitrary geometry does not exist. The purpose of this paper is, therefore, to present a solid
QLT treatment of particle drift in slab and, particularly, 2D turbulence geometry.  

The structure of the paper is as follows: Section \ref{sec:fpcder} introduces the governing  QLT equations of motions
(Sec. \ref{sec:eom}) required for the evaluation of second-order velocity cross-correlation functions (Sec. \ref{sec:ccf}). General
Fokker-Planck coefficients and drift coefficients valid for arbitrary turbulence geometry and arbitrary plasma wave
dispersion relations are presented in Sec. \ref{sec:fpc}. Drift coefficients for an electromagnetic plasma wave turbulence with slab
geometry are presented in section \ref{sec:slab}. Section \ref{sec:2d} gives a QLT particle drift coefficient for
2D turbulence geometry, presented here for the first time. The new drift coefficient is valid for an arbitrary
wavenumber dependence of the magnetic power spectrum and the turbulence decorrelation rate. Numerical calculations for a
two-component turbulence model are presented, together with the conclusions, in section \ref{sec:2dnum}. 

\section{QUASILINEAR FORMALISM}
\label{sec:fpcder}

Spatial diffusion and drift coefficients are commonly calculated from an ensemble of particle trajectories. For statistically homogeneous and
stationary fluctuations, the so-called Taylor-Green-Kubo (TGK) formula \citep[e.g.,][]{kubo1957} is often employed,
\begin{equation}
\kappa_{ij}=\int\limits_{0}^{t}d\xi<v_i(0)v_j(\xi)>,
\label{eq:tgk1}
\end{equation}
in the limit $t\to\infty$. Here, $v_i$ is the $i$th Cartesian component of the single particle velocity. The brackets $<\ldots>$ denote an ensemble average
over the relevant distribution of particles. For large coherence time $\xi$, the second-order velocity correlation function $<v_i(0)v_j(\xi)>$ must go to zero, 
and the integral in equation (\ref{eq:tgk1}) approaches a constant value for $t\to\infty$. 
 
Within the context of QLT, the transport coefficients $\kappa_{ij}$ can be written as \citep[][Eq. (12.3.25)]{schlicki2002}
\begin{equation}
\kappa_{ij}=\frac{1}{2}\int\limits_{-1}^{1}d\mu D_{X_iX_j},
\label{eq:tgk}
\end{equation}
where $\mu=v_{\|}/v$ is the pitch-angle of a particle having the velocity component $v_{\|}$ along the ordered magnetic field
$B_0$. The subscripts $X_i$ and $X_j$ denote guiding center coordinates in the $i$th and $j$th Cartesian direction, respectively,
and $D_{X_iX_j}$ denotes Fokker-Planck coefficients of the form
\begin{equation}
D_{X_iX_j}=\Re\int\limits_{0}^{\infty}d\xi
<{\dot{X_i}}(0){\dot{X_j}}^{\ast}(\xi)>.
\label{eq:fpcdef}
\end{equation}
They represent the interaction of a particle with electromagnetic fluctuations. The relation in equation (\ref{eq:fpcdef})
correspond to the TGK formula (\ref{eq:tgk1}) used earlier by \citet{bie_mat97} and \citet{for_etal74}. In QLT, however, one has to
perform an additional average with respect to $\mu$. 

To evaluate the QLT transport coefficients for particle drift in a large-scale magnetic field with superimposed electromagnetic fluctuations, the 
corresponding Fokker-Planck coefficients $D_{XY}$ and $D_{YX}$ have to be calculated. This requires the temporal variations of the quiding 
center coordinates. The particle equations of motion then enable one to calculate the corresponding second-order velocity cross-correlation functions.

\subsection{EQUATIONS OF MOTION}
\label{sec:eom}

According to \citet[][Eqs. (12.1.9d) and (12.1.9e)]{schlicki2002}, the perpendicular components of the fluctuating force fields can be written as
\begin{eqnarray}
\dot X(t) & = & -v\cos\phi(t)\sqrt{1-\mu^2}{\delta B_{\|}\over B_0}
+{c\over B_0}\left(\delta E_y+\mu{v\over c}\delta B_x\right)
\label{eq:fffx}
\\[0.2cm]
\dot Y(t) & = & -v\sin\phi(t)\sqrt{1-\mu^2}{\delta B_{\|}\over
  B_0}-{c\over B_0}\left(\delta E_x-\mu{v\over c}\delta B_y\right),
\label{eq:fffy}
\end{eqnarray}
where $\phi$ and $c$ denote the gyrophase of the particle and the speed of light, respectively. Note that the Cartesian components $\delta B_{x,y,\|}$ and 
$\delta E_{x,y,\|}$ of the fluctuating electromagnetic field are used and not the helical description, i.e., left- and right-hand polarized fields. 
For the further treatment of equations (\ref{eq:fffx}) and (\ref{eq:fffy}), a standard perturbation method is applied. To do so, it is convenient to replace in the 
Fourier transform of the irregular electromagnetic field the true particle orbit ${\bf x}(t)$ by an unperturbed particle orbit, yielding
\begin{equation}
{\bf \delta B}({\bf x},t)=\int\limits_{}^{}d^3k{\bf\delta B}({\bf k},t)e^{\imath {\bf x}(t)\cdot t}=
\sum\limits_{n=-\infty}^{+\infty}\int\limits_{}^{}d^3k {\bf \delta
  B}({\bf k},t)J_{n}(W)\exp\biggl[\imath n[\psi-\phi(t)]+\imath
  k_{\|}v_{\|}t\biggr]
\label{eq:fourier}
\end{equation}
and an analogous expression for $\delta {\bf E}$. The quantity $J_n(W)$ is a Bessel function of the first kind and order $n$. The
particle gyrophase for an unperturbed orbit is given by $\mbox{$\phi(t)=\phi_0-\Omega t$}$, where the random variable $\phi_0$
denotes the initial gyrophase of the particle. Furthermore, the abbreviation $W=k_{\perp}R_L\sqrt{1-\mu^2}$ is introduced, where
$R_L=v/\Omega$ is the Larmor radius. The relativistic gyrofrequency is given by $\Omega=qB_0/(\gamma mc)$ with $m$ being the mass and $q$ the
charge of the particle, $\gamma$ is the Lorentz factor. The angle $\psi$ results from the wavenumber representation
$k_x=k_{\perp}\cos\psi$ and $k_y=k_{\perp}\sin\psi$. With equation (\ref{eq:fourier}), the equations of motion (\ref{eq:fffx}) and
(\ref{eq:fffy}) can be manipulated to become
\begin{eqnarray}
{\dot{X}}(t)& = & {v\over B_0}\sum_{n=-\infty}^{\infty}\int\limits_{}^{}d^3 k \exp\biggl[\imath n\left[\psi-\phi(t)\right]+\imath k_{\|}v_{\|}t\biggr]
\label{eq:fffx1}
\\[0.2cm]
&&
\times\left\{-\frac{\sqrt{1-\mu^2}}{2}\Biggl[J_{n+1}(W)e^{\imath\psi}+e^{-\imath\psi}J_{n-1}(W)\Biggr]\delta B_{\|}
+{c\over v}J_n(W)\left(\delta E_y+\mu{v\over c}\delta B_x\right)\right\}
\nonumber
\\[0.4cm]
{\dot{Y}}(t) & = & {v\over B_0}\sum_{n=-\infty}^{\infty}\int\limits_{}^{}d^3 k \exp\biggl[\imath n\left[\psi-\phi(t)\right]+\imath k_{\|}v_{\|}t\biggr]
\label{eq:fffy1}
\\[0.2cm]
&&
\times\left\{\imath\frac{\sqrt{1-\mu^2}}{2}\Biggl[J_{n+1}(W)e^{\imath\psi}-e^{-\imath\psi}J_{n-1}(W)\Biggr]\delta B_{\|}
-{c\over v}J_n(W)\left(\delta E_x-\mu{v\over c}\delta B_y\right)\right\}.
\nonumber
\end{eqnarray}
For the evaluation of equation (\ref{eq:tgk}), it is convenient to consider now the nature of the electromagnetic turbulence. Here, the 
``plasma wave viewpoint'' is employed and it is assumed that the turbulence consists of a superposition of $N$ individual plasma
wave modes, i.e.,
\begin{equation}
\delta {\bf B}({\bf k},t)=\sum\limits_{j=1}^{N}\delta {\bf B}^j({\bf k})\exp(-\imath \omega_j t)\,\,;
\hspace*{0.2cm}
\delta {\bf E}({\bf k},t)=\sum\limits_{j=1}^{N}\delta {\bf E}^j({\bf k})\exp(-\imath \omega_j t).
\label{eq:super}
\end{equation}
Here, $\omega_j({\bf k})=\omega_{j,R}({\bf k})+\imath\Gamma_j({\bf  k})$ is a complex dispersion relation of wave mode $j$, where 
$\omega_{j,R}$ is the real frequency of the mode. The imaginary part, $\Gamma_j({\bf k})\leq 0$, represents dissipation of turbulent 
energy due to plasma wave damping.

Restricting the considerations to transverse ($\delta {\bf E}^j\cdot {\bf k}=0$) fluctuations and using Faraday's law, the
turbulent electric field can easily be expressed by the corresponding magnetic counterparts, yielding
\begin{equation}
\delta E_x^j=\frac{\omega_j}{ck^2}\left(\delta B_y^jk_{\|}-\delta B_{\|}^jk_y\right)
\,\,\,;\hspace*{0.1cm}
\delta E_y^j=\frac{\omega_j}{ck^2}\left(\delta B_{\|}^jk_x-\delta B_x^jk_{\|}\right).
\label{eq:fara}
\end{equation}
Furthermore, it is convenient to use the Bessel function identities 
\begin{equation} 
J_{n-1}(W)+J_{n+1}(W)= \frac{2n}{W}J_{n}(W)
\,\,\,;\hspace*{0.3cm}
J_{n-1}(W)-J_{n+1}(W) =2J_{n}^{\prime}(W),
\label{eq:besselid}
\end{equation}
where the prime denotes the derivation with respect to $W$. With equations (\ref{eq:fara}) and (\ref{eq:besselid}), 
Eqs. (\ref{eq:fffx1}) and (\ref{eq:fffy1}) can readily be rearranged to become
\begin{eqnarray}
{\dot{X}}(t)& = & -{v\over B_0}\sum\limits_{j}^{}\sum_{n=-\infty}^{\infty}\int\limits_{}^{}d^3 k \exp\biggl[\imath n\left[\psi-\phi(t)\right]+\imath (k_{\|}v_{\|}-\omega_j)t\biggr]
\label{eq:fffx2}
\\[0.2cm]
&&
\times\left\{J_{n}(W)\left[a\frac{k_x}{k_{\perp}}\delta B_{\|}^j+b\delta B_x^j\right]-\imath\sqrt{1-\mu^2}\frac{k_y}{k_{\perp}}\delta B_{\|}^jJ_{n}^{\prime}(W)\right\}
\nonumber
\\[0.4cm] 
{\dot{Y}}(t) & = & -{v\over B_0}\sum\limits_{j=1}^{N}\sum_{n=-\infty}^{\infty}\int\limits_{}^{}d^3 k \exp\biggl[\imath n\left[\psi-\phi(t)\right]+\imath (k_{\|}v_{\|}-\omega_j)t\biggr]
\label{eq:fffy2}
\\[0.2cm]
&&
\times\left\{J_{n}(W)\left[a\frac{k_y}{k_{\perp}}\delta B_{\|}^j+b\delta B_y^j\right]+\imath\sqrt{1-\mu^2}\frac{k_x}{k_{\perp}}\delta B_{\|}^jJ_{n}^{\prime}(W)\right\},
\nonumber
\end{eqnarray}
where the following complex functions have been introduced:
\begin{equation}
a=\frac{n}{W}\sqrt{1-\mu^2}-\frac{\omega_jk_{\perp}}{v\,k^2}
\,\,\,;\hspace*{0.3cm}
b=\frac{\omega_jk_{\|}}{v\,k^2}-\mu.
\label{eq:perpabini}
\end{equation}

\subsection{VELOCITY CROSS-CORRELATION FUNCTIONS}
\label{sec:ccf}

Having determined the equations of motion, one can now proceed to calculate the second-order velocity cross-correlation functions
$<{\dot{X}}(0){\dot{Y}}^{\ast}(\xi)>$ and $<{\dot{Y}}(0){\dot{X}}^{\ast}(\xi)>$ entering equation (\ref{eq:fpcdef}). The procedure for the 
calculation is relatively lengthy, but can be carried out with simple algebra. The calculations for both correlation functions are analogous, and the 
calculations are restricted to $\mbox{$<{\dot{X}}(0){\dot{Y}}^{\ast}(\xi)>$}$. Multiplication of equation (\ref{eq:fffx2}) with the conjugate of 
equation (\ref{eq:fffy2}) leads to
\begin{eqnarray}
{\dot{X}}(0){\dot{Y}}^{\ast}(\xi) & = & \frac{v^2}{B_0^2}\sum_{j}^{}\sum_{n=-\infty}^{\infty}\sum_{m=-\infty}^{\infty}\int\limits_{}^{}d^3 k\int\limits_{}^{}d^3{\overline k}\,\exp(\chi)
\label{eq:corrfun}
\\
&&
\times\Biggl\{J_{n}(W)J_{m}(\overline{W})\biggl[\frac{k_x\overline k_y}{k_{\perp}\overline k_{\perp}}a\overline a^{\ast}\cdot (\delta B_{\|}^j\delta \overline B_{\|}^{j\ast})+b\overline b^{\ast}\cdot (\delta B_{x}^j\delta \overline B_{y}^{j\ast})
\nonumber
\\
&&
\hspace*{3.5cm}
+\frac{k_x}{k_{\perp}}a\overline b^{\ast}\cdot (\delta B_{\|}^j\delta \overline B_{y}^{j\ast})+\frac{\overline k_y}{\overline k_{\perp}}\overline a^{\ast} b\cdot (\delta B_{x}^j\delta \overline B_{\|}^{j\ast})\biggr]
\nonumber
\\
&&
-\imath\sqrt{1-\mu^2}J_{m}(\overline{W})J_{n}^{\prime}(W)\frac{k_y}{k_{\perp}}
\left[\frac{\overline k_y}{\overline k_{\perp}}\overline a^{\ast} \cdot(\delta B_{\|}^j\delta \overline B_{\|}^{j\ast})+\overline b^{\ast} \cdot(\delta B_{\|}^j\delta \overline B_{y}^{j\ast})\right]
\nonumber
\\
&&
-\imath\sqrt{1-\mu^2}J_{n}(W)J_{m}^{\prime}(\overline W)\frac{\overline k_x}{\overline k_{\perp}}
\left[\frac{k_x}{k_{\perp}}a\cdot(\delta B_{\|}^j\delta \overline B_{\|}^{j\ast})+b\cdot(\delta B_{x}^j\delta\overline B_{\|}^{j\ast})\right]
\nonumber
\\
&&
-(1-\mu^2)\frac{k_y{\overline k_x}}{k_{\perp}{\overline k_{\perp}}}J_{n}^{\prime}(W)J_{m}^{\prime}(\overline{W})\cdot(\delta B_{\|}^j\delta \overline B_{\|}^{j\ast})\Biggr\}
\nonumber \end{eqnarray}
with $\chi=\imath
(n\psi-m\overline{\psi})-\imath(n-m)\phi_0-\imath({\overline
  k_{\|}}v_{\|}-\overline\omega_j^{\ast}+m\Omega)\xi$.  The bar
notation used over some quantities indicates that they have to be
evaluated for wavevector $\overline{\bf k}$ and time $\xi$.

A further simplification of (\ref{eq:corrfun}) can be achieved only if an average with respect to the random variable $\phi_0$, the initial
gyrophase of the particle, is applied. For this, the relation
\begin{equation}
\frac{1}{2\pi}\int\limits_{0}^{2\pi}d\phi_0\exp[\imath(n-m)\phi_0]=\delta_{nm}
\end{equation}
is used, where $\delta_{nm}=0$ for $n\neq m$ and unity for
 $n=m$. Furthermore, the ensemble average is applied and it is assumed
 that the Fourier components at different wave vectors are
 uncorrelated. Introducing the subscripts $\alpha$ and $\beta$ for
 Cartesian coordinates, the ensemble averages of the magnetic field
 fluctuations then read
\begin{equation}
<\delta B_{\alpha}^j\delta \overline {B}_{\beta}^{j\ast}>=<\delta B_{\alpha}^j({\bf k})\delta B_{\beta}^{j\ast}({\bf k}^{\prime})>=\delta({\bf k}-{\bf k}^{\prime})P_{\alpha\beta}^j({\bf k}).
\label{eq:mcorin}
\end{equation}
The uncorrelated state implies $\psi=\overline\psi$, $W=\overline W$ and $\omega_j=\overline\omega_j$. Equation (\ref{eq:corrfun}) can then
be manipulated to become
\begin{eqnarray}
<{\dot{X}}(0){\dot{Y}}^{\ast}(\xi)> & = & \frac{v^2}{B_0^2}\sum_{j}^{}\sum_{n=-\infty}^{\infty}\int\limits_{}^{}d^3 k\,k_{\perp}^{-2}\exp[-\imath(k_{\|}v_{\|}-\omega_j^{\ast}+n\Omega)\xi]
\label{eq:corrfun3}
\\
&&\times\Biggl\{J_{n}^2(W)\biggl[k_xk_yaa^{\ast}P_{\|\|}^j+k_{\perp}^2bb^{\ast}P_{xy}^j+k_xk_{\perp}ab^{\ast}P_{\| y}^j+k_yk_{\perp}a^{\ast}bP_{x\|}^j\biggr]
\nonumber
\\
&&-\imath\sqrt{1-\mu^2}J_{n}(W)J_{n}^{\prime}(W)\biggl[P_{\|\|}^j\left(k_y^2a^{\ast}+k_x^2a\right)+k_{\perp}k_yb^{\ast}P_{\|y}^j+k_{\perp}k_xbP_{x\|}^j\biggr]
\nonumber
\\
&&-(1-\mu^2)k_yk_x\left[J_{n}^{\prime}(W)\right]^2P_{\|\|}^j\Biggr\}
\nonumber
\end{eqnarray} 
and an analogous expression for the cross-correlation function
$<{\dot{Y}}(0){\dot{X}}^{\ast}(\xi)>$ governing the transport
parameter $\kappa_{YX}$ was derived. Both are expressed by a sum of three
individual terms, and each term is accompanied by a specific factor
through which the components of the magnetic correlation tensor
$P_{\alpha\beta}^j({\bf k},\xi)$ enter the cross-correlation functions. 

\subsection{FOKKER-PLANCK COEFFICIENTS}
\label{sec:fpc}

Having determined the velocity cross-correlation for $\kappa_{XY}$ in
the previous section, one can now proceed and evaluate the
corresponding Fokker-Planck coefficient $D_{XY}$. Upon
substituting equation (\ref{eq:corrfun3}) into (\ref{eq:fpcdef}), one obtains
\begin{equation}
D_{XY} =
\frac{v^2}{B_0^2}
\sum\limits_{j}^{}
\sum\limits_{n=-\infty}^{\infty}\Re
\int\limits_{}^{}d^3 k \frac{{\cal R}_j}{k_{\perp}^2}
\Biggl[J_n^2(W)F_{XY}^j-\imath J_n(W)J_n^{\prime}(W)G_{XY}^j-\bigl[J_n^{\prime}(W)\bigr]^2H_{XY}^j\Biggr]
\label{eq:fpcxy}
\end{equation}
with the auxiliary functions
\begin{eqnarray}
F_{XY}^j
& = & 
k_xk_yaa^{\ast}P_{\|\|}^j+k_{\perp}^2bb^{\ast}P_{xy}^j+k_xk_{\perp}ab^{\ast}P_{\| y}^j+k_yk_{\perp}a^{\ast}bP_{x\|}^j
\label{eq:fxy}
\\[0.5cm]
G_{XY}^j
& = & 
\sqrt{1-\mu^2}\biggl[P_{\|\|}^j\left(k_y^2a^{\ast}+k_x^2a\right)+k_{\perp}k_yb^{\ast}P_{\|y}^j+k_{\perp}k_xbP_{x\|}^j\biggr]
\label{eq:gxy}
\\[0.5cm]
H_{XY}^j
& = & 
(1-\mu^2)k_yk_xP_{\|\|}^j.
\label{eq:hxy}
\end{eqnarray}
The integration with respect to $\xi$ leads to the complex resonance function,
\begin{equation}
{\cal R}_{j}= \int\limits_{0}^{\infty}d\xi
\exp\left[-\imath(k_{\|}v_{\|}-\omega_{j,R}+n\Omega)\xi+\Gamma_j\xi\right]
= 
-\frac{\Gamma_j+\imath(k_{\|}v_{\|}-\omega_{j,R}+n\Omega)}{\Gamma_j^2+(k_{\|}v_{\|}-\omega_{j,R}+n\Omega)^2}
\label{eq:respwg}
\end{equation}
which describes interactions of the particles with the plasma wave
turbulence. The calculations for the Fokker-Planck coefficient
$D_{YX}$ are analogous to the calculations for $D_{XY}$ and result in
\begin{equation}
D_{YX}=\frac{v^2}{B_0^2}
\sum\limits_{j}^{}
\sum\limits_{n=-\infty}^{\infty}\Re
\int\limits_{}^{}d^3 k\frac{{\cal R}_j}{k_{\perp}^2}
\Biggl[J_n^2(W)F_{YX}^j+\imath J_n(W)J_n^{\prime}(W)G_{YX}^j-\bigl[J_n^{\prime}(W)\bigr]^2H_{YX}^j\Biggr]
\label{eq:fpcyx}
\end{equation}
with the corresponding auxiliary functions
\begin{eqnarray} F_{YX}^j
& = & 
k_xk_yaa^{\ast}P_{\|\|}^j+k_{\perp}^2bb^{\ast}P_{yx}^j+k_xk_{\perp}a^{\ast}bP_{y\|}^j+k_yk_{\perp}ab^{\ast}P_{\|x}^j \label{eq:fyx} \\[0.5cm]
G_{YX}^j
& = & 
\sqrt{1-\mu^2}\biggl[P_{\|\|}^j\left(k_x^2a^{\ast}+k_y^2a\right)+k_{\perp}k_ybP_{y\|}^j+k_{\perp}k_xb^{\ast}P_{\|x}^j\biggr]
\label{eq:gyx}
\\[0.5cm]
H_{YX}^j & = & 
\left(1-\mu^2\right)k_x k_yP_{\|\|}^j.
\label{eq:hyx}
\end{eqnarray}
Equations (\ref{eq:fpcxy}) and (\ref{eq:fpcyx}), one of the
main results of this paper and presented here in this general form
for the first time, allow one to calculate QLT drift coefficients for arbitrary turbulence geometry, where the turbulence consists of
transverse wave modes with dispersion relations depending arbitrarily
on wavevector. 

Further treatment of the auxiliary functions (\ref{eq:fxy}) to
(\ref{eq:hxy}) and (\ref{eq:fyx}) to (\ref{eq:hyx}) requires a certain
representation for the correlation tensor $P_{\alpha\beta}^j$. Different representations for $P_{\alpha\beta}^j$ will alter the underlying mathematical and physical structure of the Fokker-Planck and, therefore, the drift coefficients. Here, a representation is chosen commonly used in
the literature. Following, e.g., \citet{ler_sch01},
the nine components of $P_{\alpha\beta}^j$ can be expressed as 
\begin{equation}
P_{\alpha\beta}^j(k_{\perp},k_{\|})=A^j(k_{\perp},k_{\|})\Biggl[\delta_{\alpha\beta}-\frac{k_{\alpha} k_{\beta}}{k^2}+\imath\,\sigma^j(k_{\perp},k_{\|})\epsilon_{\alpha\beta\nu}\frac{k_{\nu}}{k}\Biggr],
\label{eq:magcorr}
\end{equation}  
where $\sigma^j$ denotes the magnetic helicity,
$\delta_{\alpha\beta}$ is Kronecker's delta and
$\epsilon_{\alpha\beta\nu}$ is the Levi-Civita tensor, $A^j$ is
the wave power spectrum. With equation (\ref{eq:magcorr}), one can now proceed to evaluate the Fokker-Planck coefficients (\ref{eq:fpcxy}) and
(\ref{eq:fpcyx}). The calculations are very laborious and result in
quite lengthy expressions. For the sake of overview, a simplification is
introduced which concerns the complex functions $a$ and $b$ given by
equation (\ref{eq:perpabini}). Their complex nature results from the
imaginary part of the wave mode dispersion relation, i.e., the
dissipation rate $\Gamma_j$. It enters $a$ and $b$ via Faraday's law
used to express the turbulent electric field components by their
magnetic counterparts. For instance, consider the first term in  equations (\ref{eq:fxy}) and (\ref{eq:fyx}). The quantity $aa^{\ast}$ can
also be written as
\begin{equation}
aa^{\ast}=\left(\frac{n}{W}\sqrt{1-\mu^2}-\frac{\omega_{j,R}k_{\perp}}{v\,k^2}\right)^2+\frac{\Gamma_jk_{\perp}^2}{v^2k^4}=a_R^2+\frac{\Gamma_jk_{\perp}^2}{v^2k^4}.
\end{equation}
Analogously, one can cast the other expressions, such as
$bb^{\ast}$ or $ab^{\ast}$, into contributions including either $\omega_{j,R}$
or $\Gamma_j$. The simplification is to neglect the contributions
given in terms of $\Gamma_j$, without the loss of insight or important
information. This reduces all equations substantially, and under the assumption that plasma wave damping is weak, this simplification seems to be reasonable. Upon substituting the appropriate components of (\ref{eq:magcorr}) into the auxiliary functions (\ref{eq:fxy}) to (\ref{eq:hxy}) and (\ref{eq:fyx}) to
(\ref{eq:hyx}), one arrives at
\begin{eqnarray}
F_{XY}^j
& = &
\frac{A^j}{k^2}\left[k_x k_y\left[(a_Rk_{\perp}-b_Rk_{\|})^2-b_R^2k^2\right]-\imath\sigma^j b_R k_{\perp}^2 k(a_Rk_{\perp}-b_Rk_{\|})\right]
\label{eq:fxyh}
\\[0.5cm]
F_{YX}^j
& = &
\frac{A^j}{k^2}\left[k_x k_y\left[(a_Rk_{\perp}-b_Rk_{\|})^2-b_R^2k^2\right]+\imath\sigma^j b_R k_{\perp}^2 k(a_Rk_{\perp}-b_Rk_{\|})\right]
\label{eq:fyxh}
\\[0.5cm]
G_{XY}^j
& = &
\frac{A^jk_{\perp}}{k^2}\sqrt{1-\mu^2}\left[k_{\perp}^2(a_Rk_{\perp}-b_Rk_{\|})-2\imath\sigma^j b_R 
k k_x k_y\right]
\label{eq:gxyh}
\\[0.5cm]
G_{YX}^j
& = &
\frac{A^jk_{\perp}}{k^2}\sqrt{1-\mu^2}\left[k_{\perp}^2(a_Rk_{\perp}-b_Rk_{\|})+2\imath\sigma^j b_R 
k k_x k_y\right]
\label{eq:gyxh}
\\[0.5cm]
H_{XY}^j
& = &H_{YX}^j
=A^j(1-\mu^2)\frac{k_x k_yk_{\perp}^2}{k^2},
\label{eq:hxyh}
\end{eqnarray} where
\begin{equation}
a_R=\frac{n}{W}\sqrt{1-\mu^2}-\frac{\omega_{j,R}k_{\perp}}{v\,k^2}
\,\,\,;\hspace*{0.3cm}
b_R=\frac{\omega_{j,R}k_{\|}}{v\,k^2}-\mu.
\end{equation}
According to equations (\ref{eq:fpcxy}) and (\ref{eq:fpcyx}), one has to
take the real part of the wavenumber integral. In view of the
resonance function (\ref{eq:respwg}) and Eqs. (\ref{eq:fxyh}) to
(\ref{eq:hxyh}), it is clear that further calculations involve both
the real and imaginary part of equation (\ref{eq:respwg}). This is not the
case for quasilinear perpendicular diffusion, where the real part of the
resonance function (\ref{eq:respwg}) is required only \citep{stawicki2004}.

In order to proceed, equations (\ref{eq:fxyh}) to (\ref{eq:hxyh}) are substituted into the coefficients (\ref{eq:fpcxy}) and (\ref{eq:fpcyx}) and the relation $a_Rk_{\perp}-b_Rk_{\|}=(k_{\|}v_{\|}-\omega_{j,R}+n\Omega)/v$ is used.
The coefficients $D_{XY}$ and $D_{YX}$ can then be expressed as
\begin{equation}
\left.
\begin{array}{lc}
\displaystyle D_{XY}
\\[0.3cm]
\displaystyle
D_{YX}
\end{array}
\right\}
={\mathsf R}\,\sin\psi\cos\psi
+
\left\{
\begin{array}{lc}
\displaystyle
-{\mathsf N}
\\[0.3cm]
\displaystyle
+{\mathsf N}
\end{array}
\right.
\end{equation}
with the operator \begin{eqnarray}
{\mathsf R} & = &
\frac{v^2}{B_0^2}
\sum\limits_{j}^{}
\sum\limits_{n=-\infty}^{\infty}
\int\limits_{}^{}d^3k\frac{\Gamma_j A^j(k_{\perp},k_{\|})k^{-2}}{\Gamma_j^2+(k_{\|}v_{\|}-\omega_{j,R}+n\Omega)^2}\Biggl\{\left[(ak_{\perp}-bk_{\|})^2-b^2k^2\right]J_n^2(W)
\label{eq:operr}
\\[0.3cm]
&&
-2\sigma^jbkk_{\perp}\sqrt{1-\mu^2}J_n(W)J_n^{\prime}(W)
-(1-\mu^2)k_{\perp}^2\bigl[J_n^{\prime}(W)\bigr]^2
\Biggr\}
\nonumber
\end{eqnarray}
representing resonant wave-particle interactions. The contribution
${\mathsf N}$ represents non-resonant behavior and reads \begin{eqnarray}
{\mathsf N} & = &\frac{v}{B_0^2}\sum\limits_{j}^{}\sum\limits_{n=-\infty}^{\infty}
\int\limits_{}^{}d^3 k\frac{A^j(k_{\perp},k_{\|})}{k^2}\frac{(k_{\|}v_{\|}-\omega_{j,R}+n\Omega)^2}{\Gamma_j^2+(k_{\|}v_{\|}-\omega_{j,R}+n\Omega)^2}
\nonumber
\\
&&\times
\Biggl\{\sigma^j\left(\frac{\omega_{j,R}k_{\|}}{v\,k^2}-\mu\right)k J_n^2(W)+\sqrt{1-\mu^2}k_{\perp}J_n(W)J_n^{\prime}(W)\Biggr\}.
\label{eq:sfn1}
\end{eqnarray}
A closer inspection of the operator (\ref{eq:operr}) results in the finding that ${\mathsf R}\,\sin\psi\cos\psi=0$, since no additional
dependences in $\psi$ are assumed. This implies vanishing resonant
wave-particle interactions. Quasilinear particle drift is then
solely characterized by the non-resonant term ${\mathsf N}$, equation
(\ref{eq:sfn1}). Upon using the relation (\ref{eq:tgk}), one obtains
\begin{equation}
\kappa_{XY}=-\kappa_{YX}=-\frac{1}{2}\int\limits_{-1}^{1}d\mu {\mathsf N}.
\label{eq:transpara2}
\end{equation}
As expected for particle drift, $\kappa_{XY}$ and
$\kappa_{YX}$ are antisymmetric. The nonresonant part,
Eq. (\ref{eq:sfn1}), allows to determine the drift coefficient (\ref{eq:transpara2}) for different turbulence
geometries. Motivated by theoretical work \citep{zankmatt1992} and
observations \citep{mattetal1990, bieetal1996}, slab and 2D turbulence
geometries are assumed here, and each is considered in turn. 

\section{DRIFT COEFFICIENT FOR SLAB GEOMETRY}
\label{sec:slab}

In slab turbulence geometry, the wavevectors are all either
parallel or antiparallel to the background magnetic field, and the
wave power spectrum can be given by
\begin{equation}
A^j=S_s^j(k_{\|})\frac{\delta(k_{\perp})}{k_{\perp}}.
\label{eq:powerspec}
\end{equation}
To calculate $\kappa_{XY}$ or, simultaneously, $\kappa_{YX}$ for slab
geometry, the Bessel functions in equation(\ref{eq:sfn1}) are considered
in the limit $W\propto k_{\perp}\to 0$. The first term in the braces
contributes only for $n=0$, since $J_n(0)=1$ for $n=0$
only. Concerning the second term, it can be shown that it vanishes
completely for $W\to 0$ and all $n$. The pitch-angle integration is
elementary and (\ref{eq:transpara2}) then yields
\begin{equation}
\kappa_{XY}^S=-\frac{\pi}{B_0^2}\sum\limits_{j=\pm 1}^{}\int\limits_{-\infty}^{\infty}dk_{\|}\sigma^j(k_{\|})S_s^j(k_{\|})k_{\|}^{-2}
\Biggl\{2\omega_{j,R}+\frac{\Gamma_j^2}{2vk_{\|}}\ln\left(\frac{\Gamma_j^2+(k_{\|}v-\omega_{j,R})^2}{\Gamma_j^2+(k_{\|}v+\omega_{j,R})^2}\right)\Biggr\},
\label{eq:kappaxy1}
\end{equation}
For further process, the wave power spectrum $S_s^j(k_{\|})=C(\nu)\lambda_s(\delta B_s^{j})^2(1+k_{\|}^2\lambda_s^2)^{-\nu}$ is employed. Here, $\lambda_s$
is the bend-over scale, $(\delta B_s^{j})^2$ is the slab variance and
$2\nu=5/3$ is the inertial range spectral index. Furthermore,
$C(\nu)=(2\sqrt{\pi})^{-1}\Gamma(\nu)/\Gamma(\nu-1/2)$, where
$\Gamma(x)$ denotes the Gamma function. Concerning the plasma wave dispersion relation, it is assumed that forward ($j=+1$) and backward ($j=-1$) propagating shear Alfv\'en waves with real frequency $\omega_{j,R}=jv_Ak_{\|}$ form the slab turbulence. For $v_A/v<< 1$, the second term in equation (\ref{eq:kappaxy1}) then becomes zero. This does not imply that the dynamical behavior of the turbulence due to dissipation is not taken into account, it is rather suppressed since the argument of the logarithmic expression becomes unity. Upon using a constant magnetic helicity, the wavenumber integration can be performed analytically and equation (\ref{eq:kappaxy1}) yields
\begin{equation}
\kappa_{XY}^S=-\kappa_{YX}^{S}=-\kappa_{0}\left[\sigma^+\left(\frac{\delta B_S^+}{\delta B_S}\right)^2-\sigma^-\left(\frac{\delta B_S^-}{\delta B_S}\right)^2\right].
\label{eq:kappaslab}
\end{equation}
Here, $\kappa_0=\pi\xi\lambda_sv_A(\delta B/B_0)^2$, where $0\leq \xi=\delta
B_s^2/\delta B^2\leq 1$ measures the fraction of
the slab contribution to the total turbulent magnetic energy, $\delta
B^2=\delta B_s^2+\delta B_{2D}^2$. The evaluation of Eq. (\ref{eq:transpara2})
for 2D turbulence geometry is presented in the next section. A closer
inspection of the slab drift coefficient (\ref{eq:kappaslab}) results
in the following findings: First, $\kappa_{XY}^S$ is entirely
determined by the magnetic helicities $\sigma^+$ and $\sigma^-$ of
forward and backward propagating wave fields, respectively. Second,
$\kappa_{XY}^S$ depends on neither the charge nor the mass of the
particle and is, therefore, independent of particle
properties. 

In view of the independence of $\kappa_{XY}^S$ of particle properties,
it is instructive to recall a standard zeroth-order drift,
the  ${\bf E}\times {\bf B}$ drift. A charged particle moves with the drift
velocity ${\bf v}_D=c({\bf E}\times {\bf B})/B^2$ if an electric force
acts normal to the background magnetic field. This drift is identical
for all charged particles and, therefore, independent of particle charge, mass and velocity. Equation (\ref{eq:kappaslab}) reveals the same
feature. An enlighting approach for a comparison is to replace
${\bf E}$ by the fluctuating field $\delta{\bf E}$. The drift velocity
${\bf v}_D$ is then a random quantity and a corresponding
velocity cross-correlation function and, therefore, drift coefficient
can be derived. Based on this, it can be shown that $\kappa_{XY}^S$ is indeed a
result of the $\delta {\bf E}\times {\bf B_0}$ drift, where the
perpendicular force results from the electric component of the
turbulence.

To obtain some more insight into the drift coefficient
(\ref{eq:kappaslab}), it is convenient to recall the definition of the
magnetic helicity $\sigma^j$. It is usually defined as
\begin{equation}
\sigma^{\pm}=\frac{(\delta B_{L}^{\pm})^2-(\delta B_{R}^{\pm})^2}{(\delta
B_{L}^{\pm})^2+(\delta B_{R}^{\pm})^2},
\end{equation}
where $\delta B_{L}$ and
$\delta B_{R}$ denote left-handed (LHP) and right-handed polarized
(RHP) field components of the fluctuations, respectively
\citep[see, e.g.,][]{schlicki2002}. Now a variety of wave fields with
different polarization states and propagation directions can be 
considered:
\begin{itemize}
\item[(1)] 
LHP wave in forward direction ($\sigma^+=+1$, $\delta
B_S^{-}=0$): equation (\ref{eq:kappaslab}) yields $\kappa_{XY}^S=-\kappa_{0}$. 
A RHP wave in backward direction ($\sigma^-=-1$,$\delta B_S^+=0$)
leads to the same result, i.e., a LHP wave in
forward direction can be replaced by a RHP wave propagating in
backward direction. If both are present, the drift becomes two times stronger, $\kappa_{XY}^S=-2\kappa_{0}$.
\item[(2)] 
RHP wave in forward direction ($\sigma^+=-1$, $\delta B_S^-=0$):
equation (\ref{eq:kappaslab}) yields $\kappa_{XY}^S=+\kappa_{0}$. 
A LHP wave in backward direction ($\sigma^-=+1$, $\delta B^+=0$) leads
to the same result. As in case (1), if both types of waves are
present, the drift becomes two times stronger, i.e., $\kappa_{XY}^S=+2\kappa_{0}$
\item[(3)]
For equal polarization states, i.e., \mbox{$\sigma^+=\sigma^-$}, one finds
$\kappa_{XY}^S=-\kappa_{0}H_c\sigma^+$ with 
\begin{equation}
H_c=\frac{(\delta B^+)^2-(\delta
B^-)^2}{(\delta B^+)^2+(\delta B^-)^2}
\label{eq:norhel}
\end{equation}
being the normalized cross
helicity \citep[][]{schlicki2002}. It represents the ratio of the
intensities of forward ($j=+1$) to backward ($j=-1$) propagating wave
fields and is sometimes also referred to as ``Alfv\'enicity''.
Obviously, $\kappa_{XY}^S$ becomes
zero for a vanishing net polarization. Furthermore, it changes sign if
predominantly LHP or RHP wave fields are present, i.e., with the
reversal of polarity.
\end{itemize}
\label{pg:case}
At a glance, the drift coefficient for slab geometry, Eq. (\ref{eq:kappaslab}), is solely determined by the magnetic helicity $\sigma^j$ and the real frequency
$\omega_{j,R}$. If one of them is neglected, no QLT particle drift
occurs in slab turbulence. However, \citet{for_etal74} and
\citet{bie_mat97} argued that FLRW governs drift in a static and purely magnetic slab turbulence. In stark contrast to this are
equations (\ref{eq:kappaxy1}) and (\ref{eq:kappaslab}). They clearly show
that particle drift in slab geometry requires turbulent electric field
components and is due to the net polarization
only, i.e., the magnetic helicity $\sigma^j$. Furthermore, the {\it
  nonresonant} FLRW limit used by \citet{for_etal74} and
\citet{bie_mat97} is based on the concept of a magnetic power spectrum
at zero wavenumber, $\delta B^2(k_{\|}=0)$. This limit can be
achieved only for static ($\Gamma_j=0$) turbulence conditions. The
{\it real} part of the resonance function (\ref{eq:respwg}) yields
then a Dirac delta distribution $\delta(k_{\|})$ required for the FLRW
limit. However, the real part represents {\it resonant}
interactions. As it is shown in Sec. \ref{sec:fpc}, the real part is
not important for QLT particle drift, and so FLRW. QLT drift in slab
geometry is a consequence of the {\it nonresonant} term, Eq. (\ref{eq:sfn1}). The latter results from the {\it imaginary} part of
equation (\ref{eq:respwg}) and can not explain the FLRW limit.

\section{DRIFT COEFFICIENT FOR 2D GEOMETRY}
\label{sec:2d}

Having derived the drift coefficient for a plasma
wave turbulence and gained the insight that fluctuating electric
fields are required for QLT drift in slab geometry, the evaluation
of the nonresonant contribution (\ref{eq:sfn1}) for 2D geometry is
presented in this section. For 2D geometry, the wavevectors are perpendicular to the mean magnetic field, and the wave power spectrum can be given by \begin{equation}
A^j(k_{\perp},k_{\|})=S_{2D}^{j}(k_{\perp})\frac{\delta(k_{\|})}{k_{\perp}}.
\label{eq:power2d}
\end{equation}
Obviously, shear Alfv\'en waves as used in the previous section for slab turbulence can not contribute, since $\omega_{j,R}=0$ in equation (\ref{eq:sfn1}) for the power spectrum (\ref{eq:power2d}). The calculations are, therefore, restricted to purely dynamical magnetic fluctuations. Since the concept of a superposition of individual wave modes does not apply anymore, the $j$-nomenclature is dropped. The only modification concerns the resonance function (\ref{eq:respwg}). To take into account the dynamical behavior of purely dynamical magnetic fluctuations, \citet{bie_etal94}
defined two models: the damping as well as the random sweeping
model. For the damping model, they suggested a
dynamical behavior of the turbulent energy being of the form
\begin{equation} <\delta B_{\alpha}({\bf k},t)\delta B_{\beta}^{\ast}({\bf k},t+\xi)>=P_{\alpha\beta}({\bf k})\exp(-\nu_c\xi),
\end{equation}
where $\nu_c$ represents decay of turbulent magnetic energy. The resonance function (\ref{eq:respwg}) has to be adapted in this respect, and the plasma wave dissipation rate $\Gamma_j$ is simply replaced by the decorrelation rate $\nu_c$. 

\begin{figure}[tbh]
\includegraphics[width=10cm]{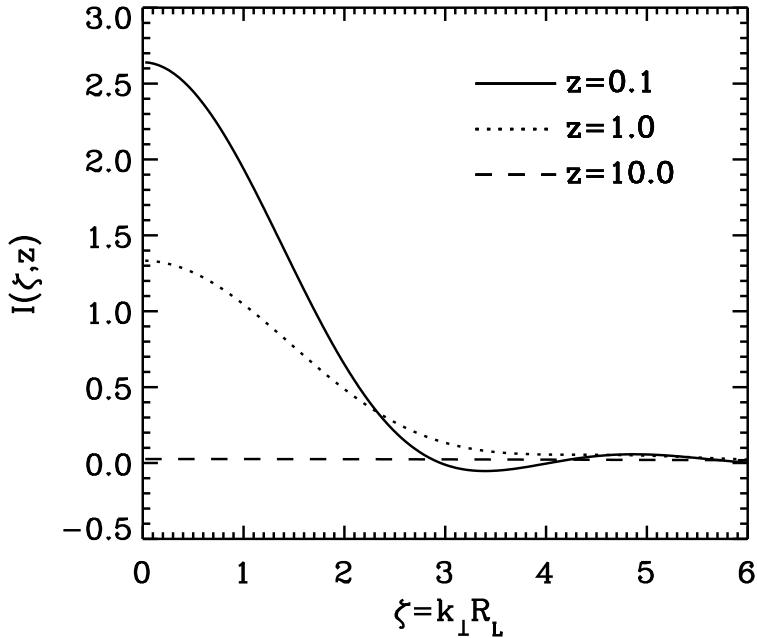}
\caption{Plot of numerical solutions to Eq. (\ref{eq:funcI}) representing
  $I(\zeta,z)$ as a function of $\zeta=k_{\perp}R_L$ for three
  different values of $z$ (see legend).
\label{fig:f1}}
\end{figure}

Upon substituting equation (\ref{eq:power2d}) into (\ref{eq:sfn1}), replacing $\Gamma_j$ by $\nu_c$ and applying the $\mu$-average, one obtains
\begin{equation}
\kappa_{XY}^{2D}=-\frac{4\pi v}{B_0^2}\sum\limits_{n=1}^{\infty}
\int\limits_{0}^{\infty}dk_{\perp}\frac{S_{2D}(k_{\perp})}{k_{\perp}}\frac{n^2\Omega^2}{\nu_c^2(k_{\perp})+n^2\Omega^2}\int\limits_{0}^{1}d\mu\sqrt{1-\mu^2}J_n(W)J_n^{\prime}(W).
\label{eq:nonres}
\end{equation}
Here, the fact was used that the first term in equation (\ref{eq:sfn1}) is an odd function in $\mu$ and, therefore, vanishes, due to the
$\mu$-average. As a consequence, the magnetic helicity $\sigma$ (not
assumed to be zero) does not influence quasilinear drift in 2D
turbulence geometry. The $\mu$-integration in equation (\ref{eq:nonres}) can be
carried out analytically and the detailed calculations are presented
in Appendix \ref{sec:app}. There, it is shown that
equation (\ref{eq:nonres}) can be manipulated to become
\begin{equation}
\kappa_{XY}^{2D}=-\kappa_{YX}^{2D}=-\frac{\pi v}{4B_0^2}R_L\int\limits_{0}^{\infty}dk_{\perp}S_{2D}(k_{\perp})I(\zeta,z)
\label{eq:kap2d}
\end{equation}
with
\begin{equation}
I(\zeta,z)=\int\limits_{0}^{\pi/2}d\theta\frac{\sin(2\theta)\sinh(2\theta z)}{\zeta^{3}\sinh(z\pi)\cos^3\theta}\biggl[2\zeta\cos\theta\cos(2\zeta\cos\theta)+(4\zeta^2\cos^2\theta-1)\sin(2\zeta\cos\theta)\biggr],
\label{eq:funcI}
\end{equation}
where the abbreviations $\zeta=k_{\perp}R_L$ and $z=\nu_c/\Omega$ are
introduced. Furthermore, $R_L=v/\Omega$ is the Larmor radius.
Equation (\ref{eq:kap2d}) is valid for a power spectrum and a
decorrelation rate varying arbitrarily in wavenumber $k_{\perp}$. It
is noteworthy that a similar expression can also be derived for
quasilinear perpendicular diffusion in 2D turbulence geometry
\citep{stawicki2004}. The integral representation (\ref{eq:funcI})
results from the $\mu$-integration and has to be evaluated for further
progress. Unfortunately, an analytical solution to this integral does
not exist, and any progress requires a numerical treatment. Figure
\ref{fig:f1} shows numerical computations of equation (\ref{eq:funcI})
as functions of $\zeta=k_{\perp}R_L$ for three different values of
$z$. For illustrative purposes, $z$ is assumed to be a constant in
$k_{\perp}$. Using small argument approximations for the hyperbolic
sine, it can be shown that Eq. (\ref{eq:funcI}) is independent of $z$
and a function of $\zeta$ only. Furthermore, note that $I(\zeta,z)$ and,
therefore, $\kappa_{XY}^{2D}$ changes sign with the reversal of a
positive to a negative particle charge state.
\\

The limit $\zeta\ll 1$, implying $R_L\ll k_{\perp}^{-1}$, leads to an
instructive, analytical solution for equation (\ref{eq:funcI}). To show this,
small argument approximations for the circular functions are used, i.e.,
$\cos(2\zeta\cos\theta)\simeq 1$ and $\sin(2\zeta\cos\theta)\simeq
2\zeta\cos\theta$ and inserted into $I(\zeta,z)$. Partial integration
then results in $I(\zeta\ll 1,z)=8/3(1+z^2)$.
Consequently, one obtains for the drift coefficient the expression
\begin{equation}
\kappa_{XY}^{2D}=-\frac{2\pi v}{3B_0^2}R_L
\int\limits_{0}^{\infty}dk_{\perp} S_{2D}(k_{\perp})\frac{(\tau_c\Omega)^2}{1+(\tau_c\Omega)^2}
\label{eq:trapafin}
\end{equation}
in the limit $R_L\ll k_{\perp}^{-1}$, where the relation
$\nu_c=\tau_c^{-1}$ has been used. Since the correlation time $\tau_c$
is still undetermined, equation (\ref{eq:trapafin}) can be considered
as being valid regardless of whether $\tau_c\Omega$ is smaller than,
larger than, or of order unity. However, in view of the restriction
$k_{\perp}R_L\ll 1$, it becomes obvious that (\ref{eq:trapafin}) is valid
only for low/intermediate particle energies if parameters are assumed
being typical for the heliosphere. An eyecatching feature of equation
(\ref{eq:trapafin}) is the term including the dimensionless product
$\tau_c\Omega$. It is formally the same as those given in equation (\ref{eq:bams}) used by \citet{for_etal74} and \citet{bie_mat97} for the FLRW limit.

\section{NUMERICAL CALCULATIONS AND CONCLUSIONS}
\label{sec:2dnum}

To demonstrate the potential and flexibility provided with the new
drift coefficients, the remaining wavenumber integration in
$\kappa_{XY}^{2D}$, Eq. (\ref{eq:kap2d}), is solved numerically and a
two-component, slab/2D turbulence is considered. Since the individual
contributions are simply additive, the total drift coefficient
$\kappa_{F}=\kappa_{XY}^S+\kappa_{XY}^{2D}$, induced by the fluctuations,
is introduced, where equations (\ref{eq:kappaslab}) and (\ref{eq:kap2d})
are used for $\kappa_{XY}^S$ and $\kappa_{XY}^{2D}$, respectively. For
the 2D component, the magnetic power spectrum
\begin{equation} S_{2D}(k_{\perp})=(1-\xi)C(\nu)\lambda_{2D}\delta B^2(1+k_{\perp}^2\lambda_{2D}^2)^{-\nu}
\label{eq:pow2d}
\end{equation} 
is employed. Here, as for the slab drift coefficient,
$C(\nu)=(2\sqrt{\pi})^{-1}\Gamma(\nu)/\Gamma(\nu-1/2)$ and $0\leq \xi=\delta
B_s^2/(\delta B_s^2+\delta B_{2D}^2)\leq 1$, where $\delta B_{2D}^2$
is the total variance of the 2D-component.
The corresponding bend-over scale is given by $\lambda_{2D}$ and
$2\nu=5/3$ is the inertial range spectral index. For the numerical
treatment, the turbulence decorrelation rate $\nu_c$ has to be
specified entering equation (\ref{eq:kap2d}) via $z=\nu_c/\Omega$. 
In analogy to \citet{bie_etal94}, $\nu_c$ is assumed to be of the form
$\nu_c=\alpha v_A k_{\perp}$ where the parameter $0\leq \alpha\leq 1$
allows adjustment of the strength of dynamical effects. The case
$\alpha=0$ represents the magnetostatic limit, $\alpha=1$ describes a
strongly dynamical magnetic turbulence.

For the numerical computations, conventional parameters being
typical for the heliosphere are used. The ratio $(\delta B/B_0)^2$ is
assumed to be $0.2$, until otherwise noted. The parameter $\alpha$ is
set to unity and the Alfv\'en speed $v_A$ is chosen to be $50$ km
s$^{-1}$. The background magnetic field $B_0$ is given by $4\cdot
10^{-5}$ Gauss. For all calculations, it is assumed that the turbulent
magnetic energy has only a small fraction in its slab component (say
$20\%$) and is dominated by the 2D turbulent energy ($80\%$), yielding
$\xi=0.2$. Until noted otherwise, it is assumed that
$\lambda_s=10\lambda_{2D}=0.03$ AU \citep{bie_etal94}. For
convenience, all Figures show absolute values of the drift
coefficients.

\begin{figure}[htb]
\includegraphics[width=10cm]{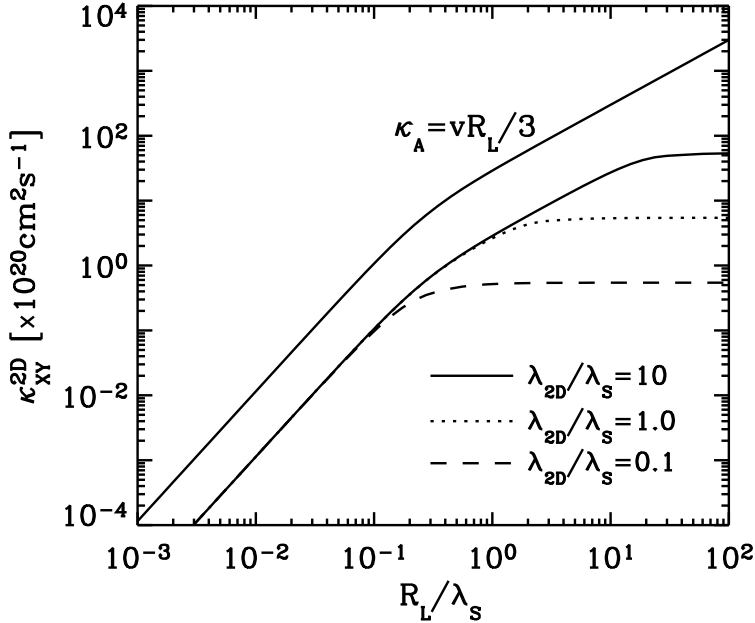}
\caption{Numerical solutions to Eq. (\ref{eq:kap2d}) representing
 $\kappa_{XY}^{2D}$ as a function of $R_L/\lambda_s$ for protons and three
 different values of the ratio $\lambda_{2D}/\lambda_{s}$ (see
 legend). The slab component is suppressed by assuming an unpolarized
 state ($\sigma^+=\sigma^-=0$), and the ratio $\delta B^2/B_0^2$ is chosen to be $0.2$. It is
 assumed that $\delta B_{2D}^2:\delta B_s^2=8:2$.
 The additional solid line visualizes the standard large-scale drift
 coefficient $\kappa_A$, Eq. (\ref{eq:largedrift}).
\label{fig:f2}}

\end{figure}

\begin{figure}[htb]
\includegraphics[width=10cm]{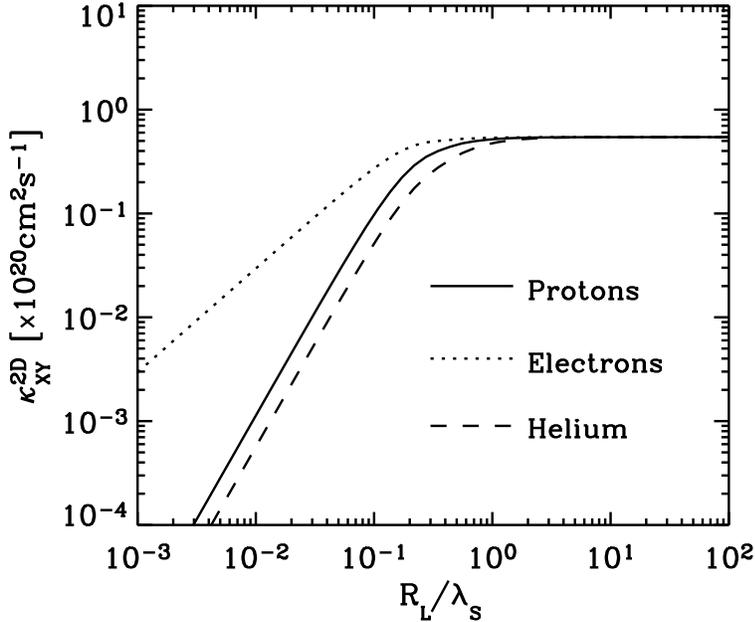}
\caption{Plot of numerical solutions to Eq. (\ref{eq:kap2d})  for $\delta 
  B^2/B_0^2=0.8$, $\lambda_{s}=10\lambda_{2D}$ and three
  different particle species: protons, electrons and
  helium.
\label{fig:f3}}
\end{figure}

Figure \ref{fig:f2} shows numerical result of $\kappa_{XY}^{2D}$,
Eq. (\ref{eq:kap2d}), as a function of the proton  Larmor radius $R_L$
normalized to the slab bend-over scale $\lambda_s$ for $\xi=0.2$
and three different values of $\lambda_{2D}/\lambda_s$ (see
  legend). The ratio $R_L/\lambda_s$ is proportional to the particle rigidity
  $R$. For the computations, it is assumed that the slab
  contribution is unpolarized, i.e., $\sigma^+=\sigma^-=0$, yielding
  $\kappa_{XY}^S=0$. The additional solid line visualizes the
  large-scale drift coefficient $\kappa_A$,
  Eq. (\ref{eq:largedrift}). A closer inspection of
  Figure \ref{fig:f2} results in the following finding: $\kappa_{XY}^{2D}$
  and $\kappa_A$ obey the same power law dependence in $R_L/\lambda_s$
  or, alternatively, rigidity $R$ for $R_L/\lambda_s\ll 1$, namely
  $R^2$. This is expected. As it is shown in Sec. \ref{sec:2d}, the
  general 2D drift coefficient (\ref{eq:kap2d}) reduces to
  equation (\ref{eq:trapafin}) for $R_L\ll k_{\perp}^{-1}$. Obviously,
  $\kappa_A$ and $\kappa_{XY}^{2D}$ then reveal the same variation in
  $R_L/\lambda_s$ or, synonymously, $R$. However, independent of the
  ratio $\lambda_{2D}/\lambda_s$, $\kappa_{XY}^{2D}$ becomes a
  constant for $R_L/\lambda_s\gg 1$, whereas $\kappa_A$ scales with
  $R$. For completeness, Figure \ref{fig:f3} illustrates numerical results of
  $\kappa_{XY}^{2D}$ for protons, electrons and helium for
  $\lambda_{2D}/\lambda_s=10$.
\\

As explained in Sec. \ref{sec:intro}, $\kappa_A$ is valid for an
unperturbed, unmodified Parker spiral only (see
Eq. (\ref{eq:largedrift}) and the comments following it), while
$\kappa_F$ describes effects due to the two-component turbulence. 
Generally, it is expected that additional (electro)magnetic turbulent
fields alter drift effects, particularly for low and intermediate particle
  energies. Whether with or without turbulence, particle motion is
  affected by curvature and gradient drift effects. The standard
  coefficient $\kappa_A$
  might therefore be considered as the limit of a more general drift
  coefficient for $\delta B\to 0$ for which $\kappa_F$ vanishes. Since
  the fluctuating fields
  are superimposed to the heliospheric background magnetic field, it
  is assumed that individual drift effects induced by $B_0$ and
  $\delta B$ are simply additive (at least in the local orthogonal   coordinate system). This results in a total drift coefficient
  $\kappa_T=\kappa_A-\kappa_F$. Note that both $\kappa_A$ as well as
  $\kappa_{XY}^{2D}$ change signs with the reversal of the particle
  charge state. According to Eq. (\ref{eq:kap2d}), but also
  equation (\ref{eq:kappaslab}) for the slab contribution, one has to take into
  account an additional minus sign, resulting in the difference of
  $\kappa_A$ and $\kappa_F$. The slab contribution is
  independent of the particle charge state and depends solely on the
  polarization of slab turbulence. In this respect, the magnetic helicity
 $\sigma^{\pm}$ of the slab component entering
  equation (\ref{eq:kappaslab}) and the normalized cross helicity $H_c$,
  Eq. (\ref{eq:norhel}), are quite uncertain parameters. A rigorous
  theory describing wavenumber and radial variations of the
  slab helicities does not exist. Usually, for the 2D component, $H_c$ is
  assumed to be zero for heliocentric distances beyond $\sim 1$ AU
  \citep[e.g.,][]{zan_etal96}, but this might probably not be the case
  \citep[see][]{mattetal2004}.

\begin{figure}[htb]
\includegraphics[width=10cm]{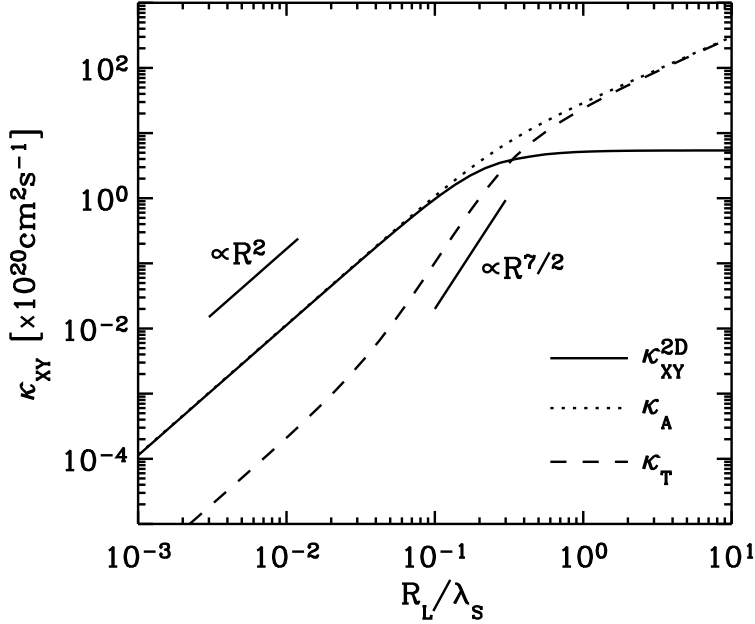}
\caption{Numerical solution of $\kappa_{XY}^{2D}$ (solid line)
 for protons and $\delta 
  B^2/B_0^2=0.8$ as well as $\lambda_{s}=10\lambda_{2D}$. The slab
  component is suppressed, $\kappa_{XY}^S=0$, by assuming an
 unpolarized state, i.e., $\sigma^+=\sigma^-=0$. The dashed curve
 represents the total drift coefficient $\kappa_T=\kappa_A-\kappa_{F}$. 
\label{fig:f4}}
\end{figure}

Figure \ref{fig:f4} shows results for $\kappa_{XY}^{2D}$ (solid
line), $\kappa_A$ (dotted line) and their difference, $\kappa_{T}$
(dashed line), for the turbulence level $\delta
B^2/B_0^2=0.8$ and an unpolarized slab contribution,
i.e., $\kappa_{XY}^S=0$. Here, protons are considered. At a glance,
the drift coefficient $\kappa_A$ is substantially reduced at low and
intermediate particle rigidities. At very low rigidities, $\kappa_T$
is almost two magnitudes smaller than the standard drift $\kappa_A$,
but reveals the same power law behavior in rigidity, i.e., $R^2$. With
further increase in $R$, $\kappa_T$ varies as $R^{7/2}$, rolling over
to $\kappa_F\propto R$ at high rigidities, where the large-scale relation
$\kappa_A$ dominates. 

As mentioned in Sec. \ref{sec:intro}, a
reduction of the amount of drift effects at low and intermediate
rigidities was suggested earlier by, e.g., \citet{potgieteretal1987} and \citet{bur90} and was then considered theoretically by
\citet{bie_mat97}. Assuming $\delta B^2/B_0^2\sim 1$,
\citet{bie_mat97} found a $R^3$ power law behavior for the scaling of
drift effects at intermediate energies. However, absolutely central to
their approach is the assumption that FLRW governs the reduction of
the drift. This was not assumed here. The FLRW limit is rather excluded, since
the {\it imaginary}, nonresonant part of the function
(\ref{eq:respwg}) governs quasilinear particle drift, and not the real
part. Furthermore, the argumentation by \citet{bie_mat97} is valid for slab geometry only. For the numerical computations given in Figure \ref{fig:f4}, the
slab component is explicitly excluded by assuming an unpolarized
state. Obviously, the reduced amount of drift effects shown in Figure
\ref{fig:f4} results rather from the presence of the 2D component than
from slab turbulence and, least at all, FLRW. Depending on the ratio $\delta
B^2/B_0^2$, the influence of magnetic perturbations
on drift effects, induced in the local orthogonal coordinate
system by curvatures and gradients of the global and unperturbed
background magnetic field, leads to a significant reduction of these
drift effects. The result $R^{7/2}$ for $\delta B^2/B_0^2=0.8$ is
relatively close to the result by \citet{bie_mat97},
i.e., $R^3$. However, Figure \ref{fig:f4} shows that $\kappa_T$
recovers to continue with $R^2$ with decreasing rigidity or ratio
$R_L/\lambda_s$. This is different when slab turbulence is polarized. 

\begin{figure}[htb]
\includegraphics[width=10cm]{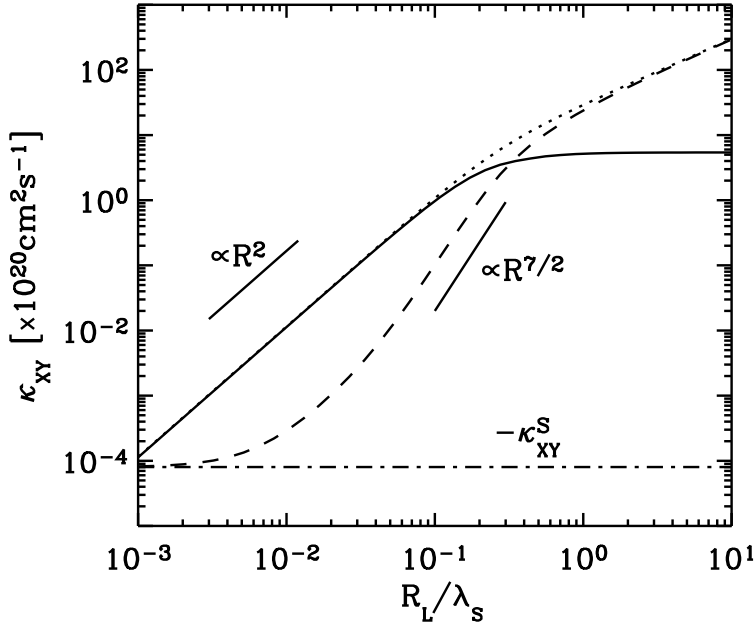}
\caption{Same as Fig. \ref{fig:f4}, but the slab component is assumed to be
 weakly right-hand polarized ($\sigma^+=\sigma^-=-0.1$). A slightly larger amount of turbulent energy in forward propagating wave fields
 is assumed, i.e. $H_c=0.1$. The dashed-dotted curve represents
 $-\kappa_{XY}^S$, Eq. (\ref{eq:kappaslab}). The dashed curve
 represents the total drift coefficient $\kappa_T=\kappa_A-\kappa_F$.
\label{fig:f5}}
\end{figure}

The influence of the slab contribution is shown in
Figure \ref{fig:f5}. Here, the same linestyle and parameters are used
as those for Figure \ref{fig:f4}, but a net polarization of the slab
component is assumed. For this, a relatively weak right-hand polarization of
forward and backward propagating wave fields is chosen,
i.e., $\sigma^+=\sigma^-=-0.1$ (see Sec. \ref{sec:slab}: item (3), page
\pageref{pg:case}). For the variances of the wave fields, the ratio
$(\delta B^-)^2:(\delta B^+)^2=9:11$ is assumed, implying for the
normalized cross helicity $H_c=0.1$. This implies a slightly larger
amount of turbulent energy in forward than in backward propagating
wave fields. A comparison of
Figures \ref{fig:f5} and \ref{fig:f4} results in the insight that the
slab drift coefficient $\kappa_{XY}^{S}$ becomes important only at
very low values of $R_L/\lambda_S$, i.e., very low particle rigidities. For this range, $\kappa_{XY}^{S}$ exceeds
$\kappa_{XY}^{2D}$ by an order of magnitude and $\kappa_{T}$ becomes a
constant for a slab component being weakly right-hand polarized. 

\begin{figure}[htb]
\includegraphics[width=10cm]{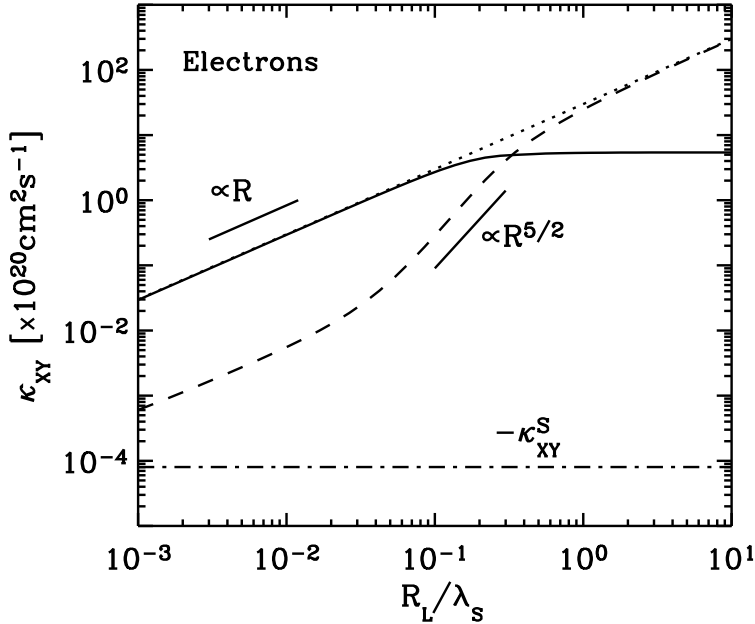}
\caption{Numerical solution of $\kappa_{XY}^{2D}$ (solid line) for the
 same parameters and linestyle as those used for the computations
 shown in Fig. \ref{fig:f5}, but here for electrons.
\label{fig:f6}}
\end{figure}

The reduction in the amount of drift effects and the change in rigidity
dependence varies for different particle species. Figure \ref{fig:f6}
shows numerical results similar to those given in Figure \ref{fig:f5},  but now for electrons. First, as expected, the 2D drift coefficient
varies as $\kappa_{XY}^{2D}\propto R$ at low and intermediate values
of $R_L/\lambda_s$, indicating the relativistic nature of the
electrons at such low rigidities. As a consequence of this,
$\kappa_{T}$ scales with $R^{5/2}$ instead of $R^{7/2}$ and rolls then
over to $\kappa_T\propto R$ at high rigidities.  
\\

The aforementioned long-standing assumption of a vanishing normalized
cross helicity $h_c$ for heliocentric distances beyond $\sim 1$ AU
\citep[e.g.,][]{zan_etal96} was made for the 2D turbulence
component only. Unfortunately, a similar treatment for the slab component
does not exist. Assuming that the same holds for slab turbulence, it
implies that $\kappa_{XY}^{S}$
vanishes in the outer heliosphere and is present only within Earth's
orbit. In this case, the 2D component dominates QLT particle drift
throughout the outer heliosphere, since it does not depend on any helicity
(see Eq. (\ref{eq:nonres}) and the comments following it). 
However, using a more advanced transport model including
cross helicity, \citet{mattetal2004} recently relaxed the assumption of
a vanishing cross helicity for 2D turbulence. Their study indicates
that cross helicity might even be present at around $10$ AU. Assuming 
that the same statement holds for slab turbulence, this would imply
that $\kappa_{XY}^{S}$ might become important not only at Earth, but also
for heliocentric distances within Jupiter's and Saturn's orbit, especially
for low rigidity particles and provided that slab turbulence is
polarized (even if weakly only). 

Another very interesting issue, which can be addressed in the context of magnetic helicity and its importance for QLT particle drift in slab geometry, 
concerns the helicity density of the large-scale HMF. \citet{bieberetal1987} have considered the topological properties of the Parker interplanetary field 
and have shown that the helicity density of the Parker field is negative north of the heliospheric current sheet and positive south of the current sheet, 
independent of the sign of the solar poloidal magnetic field. \citet{bieberetal1987} argue that the magnetic helicity of interplanetary small-scale turbulence may 
well be related to the helicity of the large-scale Parker field. This implies here that the magnetic helicity and, therefore, the polarization state of the slab 
component north of the heliospheric current sheet is opposite in sign to the polarization of the slab component south of the current sheet. 
In view of this implication, the slab drift coefficient $\kappa_{XY}^{S}$, Eq. (\ref{eq:kappaslab}), would reverse sign across the heliospheric 
current sheet, regardless of the particle charge state. Since the 2D drift coefficient $\kappa_{XY}^{2D}$, equation (\ref{eq:kap2d}), is independent of 
the magnetic helicity, only the drift coefficient for slab geometry would be affected by the change of sign across the current sheet. 
However, $\kappa_{XY}^{2D}$ changes sign with the reversal of the particle charge state (see Sec. \ref{sec:2d}). The above conclusions would add new 
elements to possible influences of the magnetic helicity (polarization) on heliospheric cosmic ray transport and their solar modulation, but the investigation 
of their impact on cosmic ray solar modulation is far beyond the scope of this paper. 

Finally, the question arises if the heliospheric transport of the so-called pick-up ions might not be affected by, at least, a polarized slab
turbulence. Effects resulting from curvature and gradient drifts and spatial diffusion are usually assumed to be small and, therefore, negligible
for this low energy particle population \citep[see][]{rucinskietal1993}.

\acknowledgments
I thank R. A. Burger, J. Minnie, H. Moraal and M. S. Potgieter for
valuable and inspiring discussions. Support by the South African National
Research Foundation (NRF) is acknowledged.

\appendix
\section{DERIVATION OF THE 2D DRIFT COEFFICIENT}
\label{sec:app}

To derive the drift coefficient $\kappa_{XY}^{2D}$,
Eq. (\ref{eq:kap2d}), for 2D turbulence geometry, the identity
 
\begin{equation}
J_n(W)J_n^{\prime}(W)=\frac{k_{\perp}R_L\sqrt{1-\mu^2}}{4n}\left[J_{n-1}^2(W)-J_{n+1}^2(W)\right]
\end{equation}
is used. Equation (\ref{eq:nonres}) can then be cast into the form
\begin{equation}
\kappa_{XY}^{2D}= -\frac{\pi vR_L}{B_0^2}
\int\limits_{0}^{\infty}dk_{\perp}S_{2D}(k_{\perp})\int\limits_{0}^{1}d\mu(1-\mu^2)\sum\limits_{n=1}^{\infty}\frac{n}{z^2+n^2}\left[J_{n-1}^2(W)-J_{n+1}^2(W)\right],
\label{eq:non2}
\end{equation}
where $z=\nu_c/\Omega$. Employing the relations \citep{gra_ryz65}
\begin{equation}
J_{n\pm 1}^2(W)=\frac{2}{\pi}(-1)^{n\pm
  1}\int\limits_{0}^{\pi/2}d\theta J_0(2W\cos\theta)\cos(2\theta [n\pm
  1]),
\end{equation}
one obtains
\begin{equation}
J_{n-1}^2(W)-J_{n+1}^2(W)=\frac{4}{\pi}(-1)^{n-1}\int\limits_{0}^{\pi/2}d\theta J_0(2W\cos\theta)\sin(2\theta n)\sin(2\theta).
\label{eq:auxbes}
\end{equation}
Upon substituting equation (\ref{eq:auxbes}) into (\ref{eq:non2}) and making use of formula (1.445.4) of \citet{gra_ryz65}, i.e.,
\begin{equation}
\sum\limits_{n=1}^{\infty}(-1)^{n-1}\frac{n\sin(2\theta n)}{z^2+n^2}=\frac{\pi}{2}\frac{\sinh(2\theta z)}{\sinh(z\pi)},
\end{equation}
equation (\ref{eq:non2}) can be manipulated to become 
\begin{equation}
\kappa_{XY}=-\frac{2\pi v}{B_0^2}R_L\int\limits_{0}^{\infty}dk_{\perp}\frac{S_{2D}(k_{\perp})}{\sinh(z\pi)}\int\limits_{0}^{\pi/2}d\theta\sin(2\theta)\sinh(2\theta z)I_{\mu}(\theta,\zeta)
\label{eq:kap2d1}
\end{equation}
with
\begin{equation}
I_{\mu}(\theta,\zeta)=\int\limits_{0}^{1}d\mu (1-\mu^2)J_0(2\zeta\cos\theta\sqrt{1-\mu^2}).
\end{equation}
Here, $W=k_{\perp}R_L\sqrt{1-\mu^2}=\zeta\sqrt{1-\mu^2}$ is used. The
$\mu$-integration can be solved analytically \citep{gra_ryz65} to obtain
\begin{eqnarray}
I_{\mu}(\theta,\zeta) & = &
\sqrt{\frac{\pi}{2}}\left[\frac{J_{1/2}(2\zeta\cos\theta)}{(2\zeta\cos\theta)^{1/2}}+\frac{J_{3/2}(2\zeta\cos\theta)}{(2\zeta\cos\theta)^{3/2}}\right]
\\
& = &
\frac{2\zeta\cos\theta\cos(2\zeta\cos\theta)+(4\zeta^2\cos^2\theta-1)\sin(2\zeta\cos\theta)}{8\zeta^3\cos^3\theta},
\nonumber
\end{eqnarray}
where $J_{\frac{1}{2}+n}(x)$ denotes spherical Bessel functions of the first
kind. The drift coefficient (\ref{eq:kap2d1}) can then be expressed as 
\begin{equation}
\kappa_{XY}=-\frac{\pi v}{4B_0^2}R_L\int\limits_{0}^{\infty}dk_{\perp}S_{2D}(k_{\perp})I(\zeta,z)
\end{equation}
with the function $I(\zeta,z)$ as defined in equation (\ref{eq:funcI}).

\end{document}